   \documentclass{aa}
   \usepackage{graphicx}
   
\begin{document}
   \title{The gas turbulence in planetary nebulae: quantification and multi-D maps from long-slit, wide-spectral range echellogram. 
 \thanks{Based on observations made with European Southern Observatory (ESO) Telescopes at La Silla (program ID 65.I-0524). 
}
 \author{ F. Sabbadin  \inst{1} \and M. Turatto \inst{2} \and S. Benetti \inst{1} \and R. Ragazzoni \inst{1} \and E. Cappellaro \inst{1} }
   \offprints{F. Sabbadin, franco.sabbadin@oapd.inaf.it}
   \institute{INAF - Astronomical Observatory of Padua, Vicolo dell'Osservatorio 5, I-35122 Padua, Italy 
   \and INAF - Astrophysical Observatory of Catania, Via S. Sofia 78, I-95123 Catania, Italy 
}}
   \date{Received January 21, 2008; accepted June 17, 2008}

   \abstract{ This methodological paper is part of a short series dedicated to the
long-standing astronomical problem of de-projecting the bi-dimensional, 
apparent morphology of a three-dimensional distribution  of gas.} {We focus on
the quantification and spatial recovery of turbulent motions in planetary nebulae (and other classes of expanding nebulae) by means of long-slit echellograms 
over a wide spectral range.}
{ We introduce some basic theoretical notions,
discuss the observational methodology, and develop an accurate
procedure disentangling all broadening components (instrumental resolution, thermal motions, turbulence, gradient of the expansion velocity, and fine 
structure of hydrogen-like ions) of the velocity profile in all spatial positions of each spectral image. This allows us to extract   
random, non-thermal motions at unprecedented accuracy, and to map them in 1-, 2- and 3-dimensions.}
{ We discuss general and specific applications of the method.  We  present the solution to practical problems in the multi-dimensional 
turbulence-analysis of a testing-planetary nebula (NGC 7009), 
using the three-step procedure (spatio-kinematics, tomography, and 3-D 
rendering) developed at the Astronomical Observatory of Padua (Italy). In addition, we  
introduce an observational paradigm valid for all spectroscopic parameters in all classes of expanding nebulae.} 
{Unsteady, chaotic motions at a local scale constitute a fundamental (although elusive) kinematical parameter of each planetary nebula, 
providing deep insights on its different shaping agents and mechanisms, and on their mutual interaction. 
The detailed study of turbulence, its stratification within a target and (possible) systematic variation among  
different sub-classes of planetary nebulae deserve long-slit, multi-position angle, wide-spectral range echellograms containing emissions 
at low-, medium-, and high-ionization,  to be analyzed 
pixel-to-pixel with a straightforward and versatile methodology, extracting all the physical information (flux, kinematics, electron temperature and 
density, ionic and chemical abundances, etc.) stored in each  frame at best.} 
\titlerunning{The gas turbulence in planetary nebulae }
\maketitle
%
\section{Introduction} 
The presence of circumstellar and interstellar gas in expansion is the typical signature of instability phases in 
stellar evolution, characterized by a large, prolonged  mass-loss rate (planetary and  symbiotic star nebulae, shells around Wolf-Rayet and 
luminous blue variable [LBV] stars), or even  explosive events (nova and supernova remnants). 
The mass-loss of evolved stars occupies a strategic 
ground between stellar and interstellar physics by raising fundamental astrophysical problems (e. g., origin and structure of winds, 
formation and evolution of dust, synthesis of complex molecules), by playing a decisive role in the final stages of stellar evolution, 
and because it  
is crucial for the galactic enrichment in light and heavy elements. 

In particular, the late evolution of low-to-intermediate 
mass stars (1.0 M$_{\odot}$$\le$M$_*$$\le$8.0 M$_{\odot}$) 
is marked by the planetary nebula (PN) metamorphosis:
an asymptotic giant branch (AGB) star gently pushes out surface layers in a florilegium 
of forms, crosses the HR diagram, and reaches the white dwarf regime.  \footnote{The foregoing sentences are taken from 
\cite{tur02} (2002), and \cite{sab06} (2006), 
of whom the present paper is the ideal sequel.}

According to current hydrodynamical and radiation-hydrodynamical simulations 
 (\cite{mel94} 1994; \cite{mar01} 2001; \cite{vil02} 2002; \cite{per04a} 2004a, 2004b; \cite{scho05} 2005), 
the PN evolution  is driven by its central star through (a) the AGB mass-loss history; 
and (b)  the variation of ionizing flux and wind power in the post-AGB phase (without excluding the possible contribution 
of other factors, like magnetic fields and binarity of the central star; 
\cite{sok98} 1998; \cite{gar99} 1999; \cite{mas99} 1999; \cite{fra99} 1999). 

At present, it is still unclear the relative importance of different 
shaping agents, their mutual influence, 
and how they affect the (global and small-scale) structure and kinematics of the ejected gas. 
Leaving simplistic ``modern views'' of PN evolution out of consideration, we remark that updated   
1-D radiation-hydrodynamical evolutionary models  
(\cite{per04a} 2004b; \cite{scho05} 2005) 
still partly fail  in reproducing the 
positive expansion velocity gradient of the main shell (the Wilson's law; 
\cite{wil50} 1950;  \cite{wed68} 1968), and the gas distribution observed in a representative sample of  carefully 
de-projected ``true'' PNe (\cite{sab06} 2006 and references therein). 
Such model-PN vs. true-PN discrepancies suggest that the former tend to systematically overestimate the relative 
importance of wind interaction (as supported by a recent revision downward of mass-loss rates in PN central stars, due to 
clumping of the fast stellar wind; 
\cite{kud06} 2006; \cite{pri07} 2007).

An even deeper uncertainty surrounds the turbulence of expanding gas, a fundamental parameter measuring local, random, unsteady 
motions due to non-linear energy transfer. 
On the one hand, 2-D hydrodynamical simulations of wind interaction  
(\cite{fem94} 1994; \cite{vis94} 1994; \cite{mel97} 1997; \cite{dwa98} 1998; \cite{gar99} 1999, 2006) 
predict stratified turbulent motions peaking within shocked 
regions  and/or ionization fronts characterized by the growth of Kelvin-Helmholtz, Vishniac  or Rayleigh-Taylor 
instabilities. 
On the other hand, practical determinations based on high-resolution spectra 
(\cite{gue98} 1998; \cite{ack02} 2002; \cite{gaz03} 2003, 2006; \cite{gz07} 2003, 2007; and \cite{med06} 2006)  
provide a unique, average turbulence-value for each nebula, ranging from 0 km s$^{-1}$ to 20 km s$^{-1}$. 

We aim at quantifying and mapping turbulent 
motions in PNe by means of long-slit, wide spectral range, very-high spectral and 
spatial resolution echellograms at several position angles (PA), reduced and analyzed according to spatio-kinematical, tomographic and 3-D 
recovery methods developed at the Astronomical Observatory of Padua 
(\cite{sab00} 2000, 2006, 
and references therein).

This introductory, methodological  paper (a) dissects all  broadening agents of a velocity profile; (b) describes an original procedure 
mapping 
turbulent motions in 1-, 2-, and 3-dimensions; (c) illustrates the solution to practical problems of turbulence-determination in a testing-PN; and 
(d) introduces an observational   
paradigm valid for all types of expanding nebulae. 

In a forthcoming paper, we (i) will present multi-dimensional turbulence maps for a representative sample of PNe in different 
evolutionary phases; and (ii) will compare these observational results with expectations from hydrodynamical simulations and theoretical models.  
\section{Basic considerations}
 A PN consists of (large- and small-scale) interconnected sub-systems with 
different morphology, physical conditions, and kinematics. 
Each elementary volume within the radial slice of nebula selected by 
a spectrograph slit is characterized by  
\begin{description}
\item[1)] a regular expansion, and  
\item[2)] random (i. e., thermal and turbulent) motions. 
\end{description}
Thanks to the large stratification of the radiation and the kinematics, each ionic 
species generates the characteristic bowed-shaped spectral image sketched in Fig. 1\label{Figure 1}. 
\begin{figure}
   \centering \includegraphics[width=9cm,height=8cm]{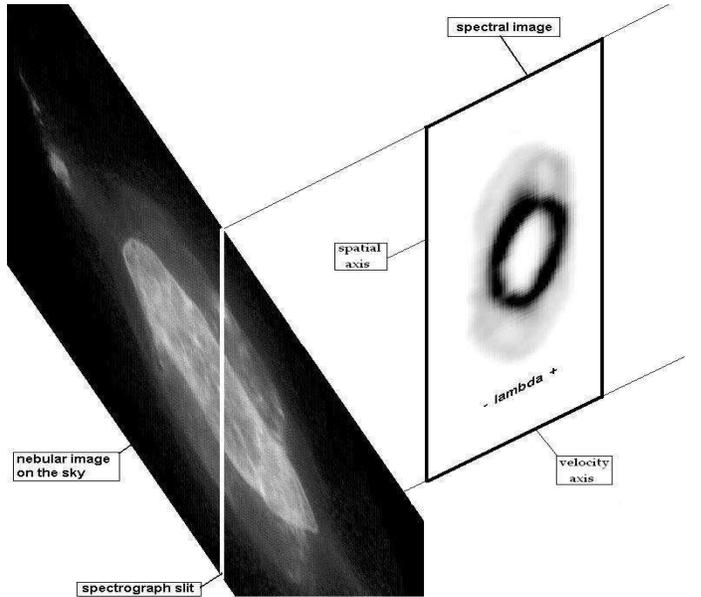}
   \caption{Nebular projection on the sky vs. high-dispersion spectral image. The radial slice of the Saturn Nebula intercepted by a spectrograph 
slit crossing the central star generates a series of high-resolution spectral images; each spectral image contains  
kinematical and spatial information on the parent ionic species. } 
\label{Figure 1}
\end{figure}

Two useful and complementary reference lines of a spectral image 
(\cite{sab00} 2000, 2004, 
and references therein) are: 
\begin{description}
\item [-] the {\bf central-star-pixel-line} (cspl), common to all PA, representing the emission of the matter projected at the apparent position of the central 
star, whose motion is purely radial;   
\item [-] the {\bf zero-velocity-pixel-column} (zvpc), giving the spatial intensity profile of the tangentially 
moving gas at the systemic radial velocity. 
\end{description}
 
As illustrated in Fig. 2\label{Figure 2}, the cspl of a spectral image contains the kinematical information:   
the intrinsic velocity profile of each nebular sub-system, convolved with a series of broadening agents (instrumental resolution, thermal motions, 
turbulence, and fine-structure of hydrogen-like ions), produces an observed, ``Gaussianized''  velocity profile, whose 
peak position and full-width-at-half-maximum (FWHM) of the   
blue- and red-shifted components give the radial velocity and the velocity spread of approaching and receding layers, respectively. 

Analogously, the zvpc of a spectral image contains the spatial information: the intrinsic spatial  profile of each nebular sub-system, convolved 
for seeing+guiding effects, produces a ``Gaussianized''  spatial profile, whose 
peak separation and FWHM give the angular distance and the thickness, respectively,  of the corresponding 
ionic layers in the plane of the sky.

\begin{figure}
   \centering \includegraphics[width=9cm]{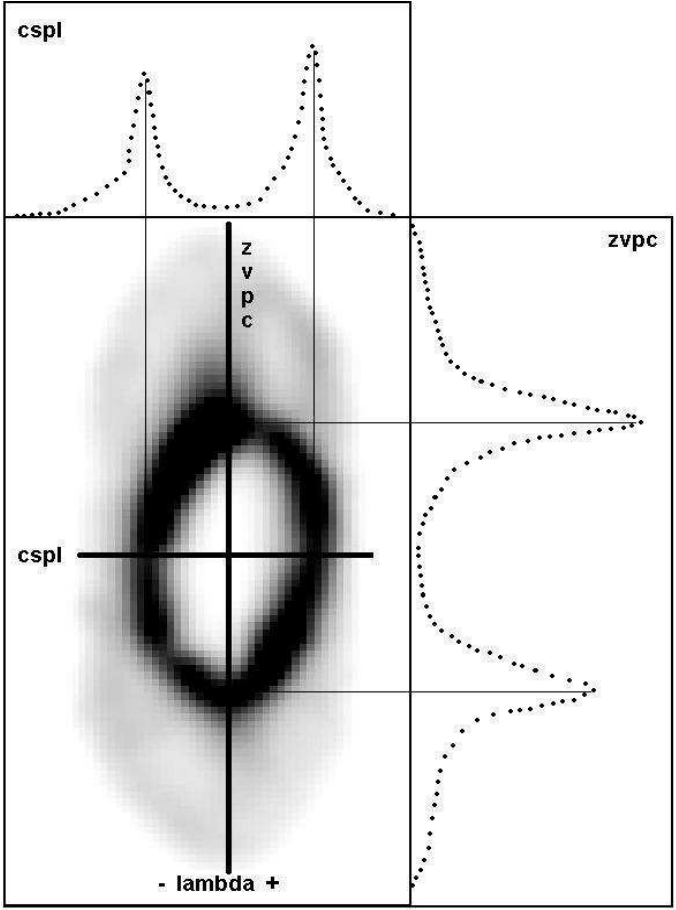}
   \caption{Anatomy of a high-resolution spectral image, showing the location  and flux distribution of both the central-star-pixel-line 
(cspl)  and  the zero-velocity-pixel-column (zvpc). More details are in the text.} 
\label{Figure 2}
\end{figure}

Let us consider the cspl, and focus on the FWHM of blue- and red-shifted components ($W_{total}$(blue) and 
$W_{total}$(red), respectively). For a typical  nebular sub-system, the ``Gaussianization'' of the cspl is very effective, and the value of $W_{total}$ is 
the sum (in quadrature) of different broadening contributions, namely:  
\begin{description}
\item [(A)] {\bf The gradient of the expansion velocity}, corresponding to the range in radial velocity of each    
emitting layer, which is usually assimilable to a Gaussian profile with $W_{grad}$.  The value of $W_{grad}$ changes from one ion to another, and can be estimated 
from the  expansion law and 
the spatio-kinematical reconstruction. A common case concerns an ellipsoidal PN seen at random orientations and expanding with  
$V_{\rm exp}$=A$\times$r''; we have   
$W_{grad}$=A$\times$$\Delta$r'', where 
the angular thickness $\Delta$r'' is given by the zvpc profile. 
Typical $W_{grad}$ values in PN main-shells are 10 to 15 km s$^{-1}$ for H I, 7 to 10 km s$^{-1}$ for He I and He II, and 4 
to 6 km s$^{-1}$ for ionic species of heavier elements. 

\item [(B)] {\bf  Instrumental resolution}, corresponding to a Gaussian profile with FWHM ($W_{instr}$), is  
given by the spectral resolution ($\lambda$/$\Delta\lambda$) of the spectrograph; a typical value of 
$\lambda$/$\Delta\lambda$=50\,000 provides W$_{instr}$=6.0 km s$^{-1}$.

\item [(C)] {\bf Thermal motions}, generating a Gaussian distribution with 
$W_{ther}$=(8$\times$k$\times$$T_{\rm e}$$\times$ln 2/${\rm m}$)$^{0.5}$= 0.214$\times$$T_{\rm e}$$^{0.5}\times$${\rm m}$$^{-0.5}$ km s$^{-1}$, 
where $T_{\rm e}$ is the local electron temperature, and m is the 
atomic weight of the element. Thus, for $T_{\rm e}$=10$^4$ K, $W_{ther}$  amounts to 21.4 km s$^{-1}$ for hydrogen, 10.7 
km s$^{-1}$ for helium, 5.8 km s$^{-1}$ for nitrogen, and 5.4 km s$^{-1}$ for oxygen. 

 \item [(D)] {\bf Turbulence}, i. e., non-thermal, chaotic motions at a local scale, is characterized by $W_{turb}$.\footnote{We 
``assume'' a Gaussian probability distribution for the turbulence velocity, although some authors suggest a non-Gaussian behavior 
for vorticity, energy dissipation and passive scalars (for a review, see 
\cite{elm04} 2004
).} Due to the intrinsically  
dissipative nature of turbulence (and in agreement with hydrodynamical simulations; 
e. g., \cite{mel97} 1997; \cite{dwa98} 1998; \cite{gar99} 1999), 
$W_{turb}$ varies 
across the nebula, peaking in shocked layers and ionization fronts.  Thus,  different ionic species are affected by different turbulent 
motions, and, on the analogy of the radiation and the kinematics, we can introduce the concept  of ``stratification of the turbulence''. 
This means that a thorough 
investigation of the gas turbulence needs spectra containing emissions at high-, medium-, and low-ionization.
\item [(E)] {\bf  The fine-structure of hydrogen-like ions}, whose levels with the same principal 
quantum number have small energy differences due to the spin-orbit coupling. We have performed a detailed analysis for Case B of 
\cite{bak38} (1938), and a {\it l}-levels population proportional to their statistical weights (\cite{sto95} 1995), 
obtaining  $W_{fs}$=7.7($\pm$0.3) km s$^{-1}$ and 5.9($\pm$0.3) km s$^{-1}$ for the seven fine-structure components of H I H$\alpha$ and 
H$\beta$, respectively (also see 
\cite{cle99} 1999); 
10.1($\pm$0.3) km s$^{-1}$ for $\lambda$ 4686 $\rm\AA\/$ of He II (thirteen fine-structure components); 4.2($\pm$0.3) km s$^{-1}$  
for  $\lambda$ 6560 $\rm\AA\/$ of 
He II (nineteen fine-structure components);  and 3.0($\pm$0.3) km s$^{-1}$ for $\lambda$ 5876 $\rm\AA\/$ of He I (six fine-structure components). 
Moreover, the fine-structure profile of He II $\lambda$ 4686 $\rm\AA\/$ presents a weak - although showy - blue-shifted tail, and the one of 
He I $\lambda$ 5876 $\rm\AA\/$ a modest red-shifted tail. 
The values of $W_{fs}$ are nearly independent on $T_{\rm e}$ for 5000 K$\le$$T_{\rm e}$$\le$20000 K, and on the electron density ($N_{\rm e}$) for 
10$^2$ cm$^{-3}\le$$N_{\rm e}$$\le$10$^5$ cm$^{-3}$.
\end{description}
For a typical nebular sub-system, the net turbulence-broadening in each ionic species is 
\begin{equation}
{W_{turb}=(W_{total}^2 - W_{grad}^2 - W_{ther}^2 - W_{instr}^2 - W_{fs}^2)^{0.5}}.  
\end{equation}

The application of Eq. (1) to the cspl of emissions at low-, mean-, and high-ionization, combined with  
de-projection criteria introduced by 
\cite{sab05} (2005), 
provides the {\bf overall 1-D turbulence-map} of the 
nebula.

To be noticed: beyond the cspl, Eq. (1) applies to each pixel-line along the dispersion, thus disentangling 
all broadening agents of the velocity profile at all ``latitudes'' of a spectral image. A different-latitude-pixel-line is identified 
by its angular distance (d$\arcsec$) from the cspl;  on the  analogy of the latter, in the following it will be named ``dlpl''. 

Thus,
\begin{description}
\item[ I -] The application of Eq. (1) to all dlpl of a spectral image allows us 
to map the turbulence of the parent   
ion within the entire de-projected nebular slice covered by a spectrograph slit ({\bf ionic 2-D turbulence-map}); 
\item[ II -] The combination of all ionic 2-D turbulence-maps at a given PA provides the  
{\bf overall 2-D turbulence-map} at the corresponding PA; and   
\item[ III -] By assembling 
overall 2-D turbulence-maps at all observed PA (for example, through the 3-D rendering procedure introduced 
by Ragazzoni et al. 2001), we infer the 
{\bf overall 3-D turbulence-map}  of the entire nebula. 
\end{description}
Summing up, the determination of the local turbulence in 1-, 2-, and 3-dimensions passes through the constraining of  
sinergistic broadening 
effects of both the instrumentation and nebular gas parameters (the latter include atomic properties, physical conditions, kinematics and ionization structure). 
This implies that {\it an adequate instrumental spectral resolution ($\lambda$/$\Delta\lambda\ge$50\,000)
is the ``conditio sine qua non'' for a quantitative study of turbulent motions in PNe, to be combined with 
a detailed spatio-kinematical reconstruction at high-, medium-, and low-ionization}.  

In addition, large, extended nebular sub-systems (e. g., optically-thin outer shells) with a broad, flat-topped intrinsic velocity profile departing 
from a Gaussian distribution, provide less reliable turbulence-values than localized, compact sub-systems with a sharp intrinsic velocity profile (e. g., 
the main shell, condensations and knots, etc.). 
Last, recombination lines of H I, He I and He II are less reliable turbulence-diagnostics than collisionally-excited lines of heavier elements.
\begin{figure*}
   \centering \includegraphics[width=14cm]{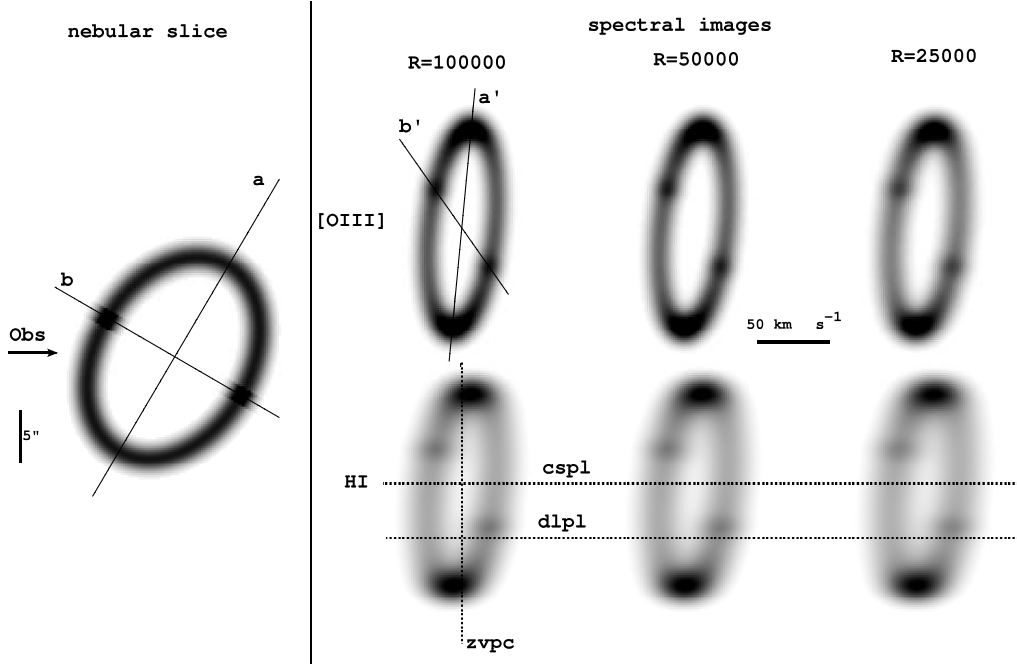}
   \caption{A model-nebular slice and the corresponding spectral image in the collisionally-excited line $\lambda$ 5007 \rm\AA\/ of [O III] and in the  
recombination line 
$\lambda$ 6563 \rm\AA\/ of H I for different values of the 
instrumental spectral resolution. Left: model-nebular slice selected by a spectrograph slit centered on the apparent major axis of 
a tilted spheroid with 
polar diameter=20\arcsec, 
axial ratio=1.4, radial thickness(FWHM)=2.0\arcsec \, and $V_{\rm exp}$(km s$^{-1}$)=3.0$\times$r\arcsec. 
Right: corresponding [O III] (upper) and H I (lower) spectral 
images for three values of the instrumental resolution R=$\lambda$/$\Delta\lambda$. } 
\label{Figure 3}
\end{figure*}

All this is illustrated in Fig. 3\label{Figure 3}, showing [O III] and H I spectral images of a random seen, 
regularly expanding ($V_{\rm exp}$$\propto$r) 
model-spheroid 
denser at the 
equator than at 
the poles, covered along 
the apparent major axis. We adopt three representative 
values of the instrumental 
resolution ($\lambda$/$\Delta\lambda$=100\,000, 50\,000, and 25\,000). Moreover, we assume $T_{\rm e}$=10$^4$ K, W$_{turb}$=10 
km s$^{-1}$, seeing=1.0\arcsec, and instrumental spatial and spectral scales of 0.20\arcsec  px$^{-1}$ and 1.5 km s$^{-1}$ px$^{-1}$, 
respectively. Different acronyms used in the current 
discussion (cspl, dlpl, zvpc etc.) 
are indicated in Fig. 3\label{Figure 3}, together with straight lines (a and b) containing polar and 
equatorial axes of the nebular slice, and their projections (a' and b') on a spectral 
image. According to the second de-projection rule introduced by 
\cite{sab05} (2005),  
x-y coordinates  of the nebular slice in Fig. 3 are related to x'-y' coordinates of the spectral image by: 
\begin{equation}
y=y'
\end{equation}
\begin{equation}
x=x'\times\vert a''\times b''\vert, 
\end{equation}
where {\it a''} and {\it b''} are the slopes of straight lines a' and b'. 

The foregoing analysis stresses a fundamental difference between imaging and spectroscopy: the former contains the 
bi-dimensional projection of the nebula on the sky, whose accuracy depends only on ``external'' factors (i. e., seeing and Airy disk 
for ground-based and orbiting telescopes, respectively), whereas the latter provides the pixel-to-pixel flux and velocity in different 
ions, and allows 
the nebular de-projection, whose accuracy depends on both ``external'' (instrumental resolution, seeing) and ``intrinsic'' (physical conditions, 
ionization, expansion, and turbulence) contributions. 

The odd thing is that nebular spectra - storing much more physical information than direct imagery - are commonly integrated along 
the spectrograph slit, and used to infer the mean, average value of flux, electron temperature, electron density, ionic and chemical abundances, 
velocity, turbulence, etc. (thus, losing their precious spatial information). 


\section{The solution to practical problems in the multi-dimensional turbulence-analysis}
Here we solve problems related to the quantification and distribution of turbulent 
motions in real PNe. To this end, we apply the procedure developed in Sect. 2 to a representative testing-target contained in 
the multi-PA, long-slit, wide spectral range, high-resolution 
spectroscopic survey of PNe in both hemispheres carried out with ESO NTT+EMMI (spectral range $\lambda\lambda$ 3900-8000 $\rm\AA\/$; 
$\lambda$/$\Delta\lambda$=60\,000) and Telescopio Nazionale Galileo (TNG)+SARG (spectral range $\lambda\lambda$ 4600-7900 $\rm\AA\/$; 
$\lambda$/$\Delta\lambda$=114\,000). 

First, a basic observative note: our long-slit spectra cover the entire instrumental spectral range 
(80 and 54 orders for ESO NTT+EMMI and TNG+SARG, respectively), whereas long-slit echellograms of extended PNe  
are usually secured through 
an interference filter isolating a single order, to avoid the superposition of echelle orders.
We stress that the superposition of echelle orders is unimportant for an emission-line object like a PN, the actual limit 
being the effective superposition of spectral images in adjacent orders. Thus, a single long-slit, wide-spectral range echellogram 
provides the complete spectral image in a number of emissions belonging to main ionic species, and including $T_{\rm e}$ and $N_{\rm e}$ 
diagnostics (\cite{tur02} 2002; \cite{ben03} 2003; and \cite{sab04} 2000, 2004, 2005, 2006).  
This is illustrated in Fig. 4\label{Figure 4}, showing the  $\lambda\lambda$ 3900-8000 $\rm\AA\/$ long-slit ESO NNT+EMMI 
rich-spectrum of \object{NGC 3918} (PNG 294.6+04.7, 
\cite{ack92} 1992). 

\begin{figure*}
   \centering \includegraphics[width=17cm]{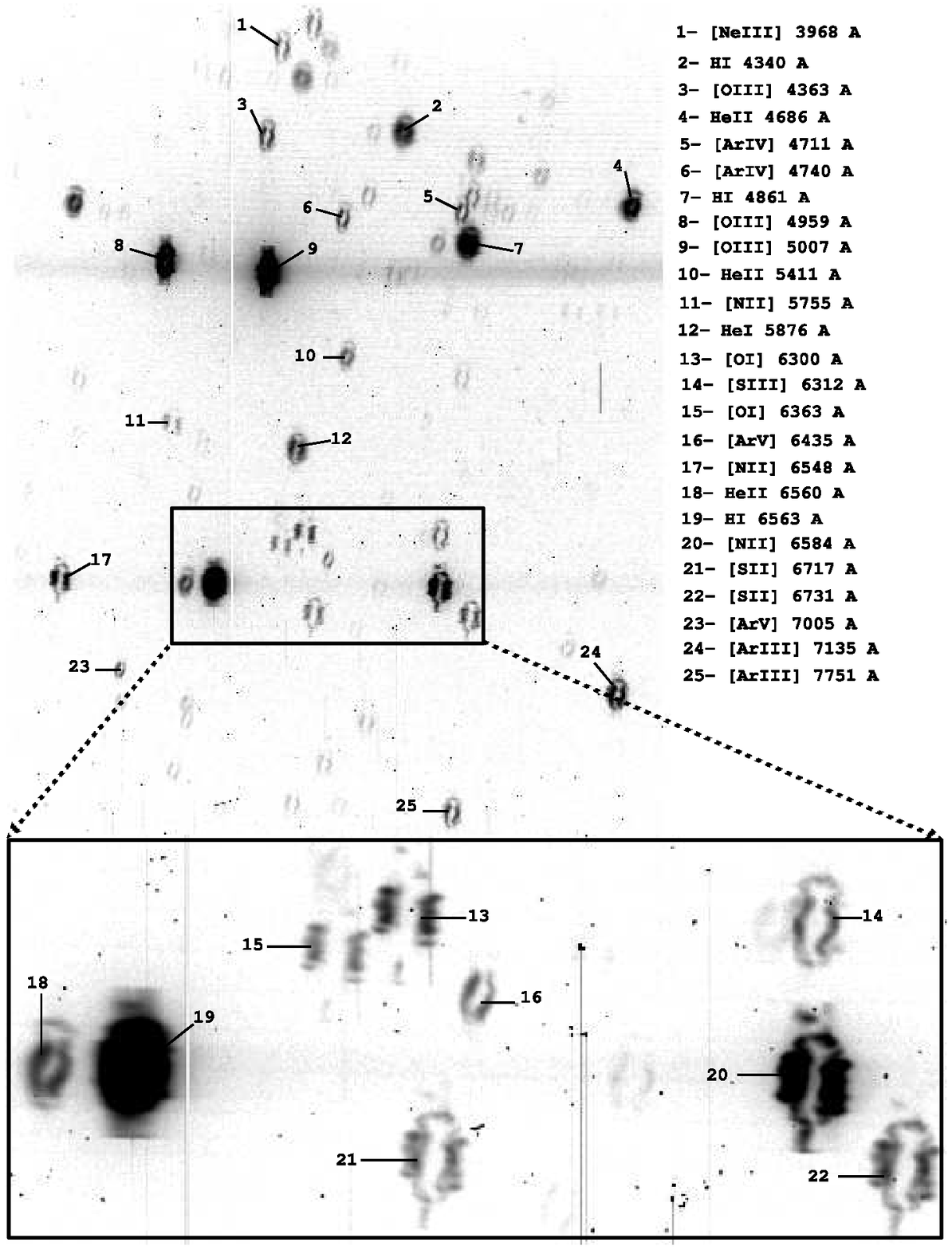}
   \caption{The upper panel shows the main part of a 
$\lambda\lambda$ 3900-8000 $\rm\AA\/$, long-slit (30$\arcsec$) ESO NTT+EMMI spectrum (echelle order separation=4$\arcsec$ (blue) to 
13$\arcsec$ (red)) of NGC 3918 at 
PA=101$\degr$ (apparent nebular size $\simeq$27$\arcsec$). 
The brightest emissions are identified.  
The lower part of the figure is an enlargement of the red spectral region; it enhances the richness of physical 
information stored in each echellogram covering 80 orders (to be compared with the single-order output provided by 
the usual observational procedure inserting an interference filter).} 
\label{Figure 4}
\end{figure*}

The order-by-order reduction procedure for long-slit, wide-spectral range echellograms uses IRAF \footnote{IRAF is distributed by the National Optical 
Astronomy Observatory, which is operated by the Association of Universities for Research in Astronomy, 
Inc., under cooperative agreement with the National Science Foundation.}  and IDL packages, follows the standard method for 
bi-dimensional spectra 
(bias, CCD-flat, distortion correction, and wavelength and flux calibrations), and is fully described by 
\cite{tur02} (2002). 

We select the bright PN \object{NGC 7009} (the Saturn Nebula, PNG 037.7-34.5, 
\cite{ack92} 1992) - observed at twelve equally-spaced PA with 
ESO NTT+EMMI (slit-length=60$\arcsec$) -  as a test-case for the solution of practical problems in the multi-dimensional turbulence-analysis.

\subsection{1-D and 2-D turbulence-analyses of NGC 7009 close to its apparent minor axis}

The Saturn Nebula is an extremely complex object, consisting of several interconnected 
sub-systems (including optically-thick caps and ansae, and high-excitation 
polar streams) with different morphology, physical conditions and kinematics. However, nebular regions projected along and close 
to the apparent minor axis are optically thin to the 
UV radiation of the central star, and their  spatio-kinematical structure is relatively simple and regular 
(\cite{sab04} 2004). 

Spectral images of main ionic 
species present in NGC 7009 at PA=124$\degr$ (i. e. close to its apparent minor axis) are shown in Fig. 5\label{Figure 5}, in order  
(left to right) of decreasing ionization potential, IP. 
They mark the spatio-kinematical signature 
in both the main shell and the outer shell of the Saturn Nebula. According to 
\cite{sab04} (2004),  
\begin{description}
\item[(1) ] the former expands at  
$V_{\rm exp}$(main shell)=4.0($\pm$0.3)$\times$r$\arcsec$ km s$^{-1}$; 
\item[(2) ]the latter is slightly blue-shifted (by 3($\pm$1) 
km s$^{-1}$), and moves at $V_{\rm exp}$(outer shell)=3.15($\pm$0.3)$\times$r$\arcsec$ km s$^{-1}$; and 
\item[(3) ] de-projected tomographic 
maps in different ions are nearly circular rings.  
\end{description}
Only medium- to high-excitation emissions are contained in Fig. 5\label{Figure 5}, NGC 7009 being optically-thin along and close to the 
apparent minor axis, with a modest concentration of low-ionization species (O$^0$, O$^+$, S$^+$, N$^+$ etc.). 

We start with the cspl (common to all PA, and providing the 1-D turbulence-map), and then extend the analysis to the whole spectral image.

\subsubsection{The 1-D turbulence-profile from the cspl}
Overall cspl flux-profiles in NGC 7009 at PA=124$\degr$ are shown in Fig. 6. Besides 
the large broadening of H I and He 
II emissions, and  the faint blue-shifted (red-shifted) tail characterizing the fine-structure profile of He II $\lambda$ 4686 $\rm\AA\/$ 
(He I $\lambda$ 5876 $\rm\AA\/$), Figs. 5 and 6 \label{Figure 5,Figure 6} evidence a strong line-asymmetry due to the slightly blue-shifted outer shell. 

Thus, we  extract the net spectral contribution in both shells through a multi-Gaussian analysis (SPLOT routine 
of the IRAF package), as illustrated in Fig. 7\label{Figure 7} 
for [O III]  $\lambda$ 5007 $\rm\AA\/$. Such a detailed deconvolution highlights all kinematical differences between nebular sub-systems, and in 
particular, the broader velocity profile of the outer shell ($W_{total}$(blue)=21.3 km s$^{-1}$, 
and $W_{total}$(red)=20.0 km s$^{-1}$ in Fig. 7, to be compared with the corresponding value of 10.0 km s$^{-1}$ for both blue- and 
red-shifted [O III] components of the 
main shell). 

We analyze these two nebular sub-systems separately.

\begin{figure*}
   \centering \includegraphics[width=15cm]{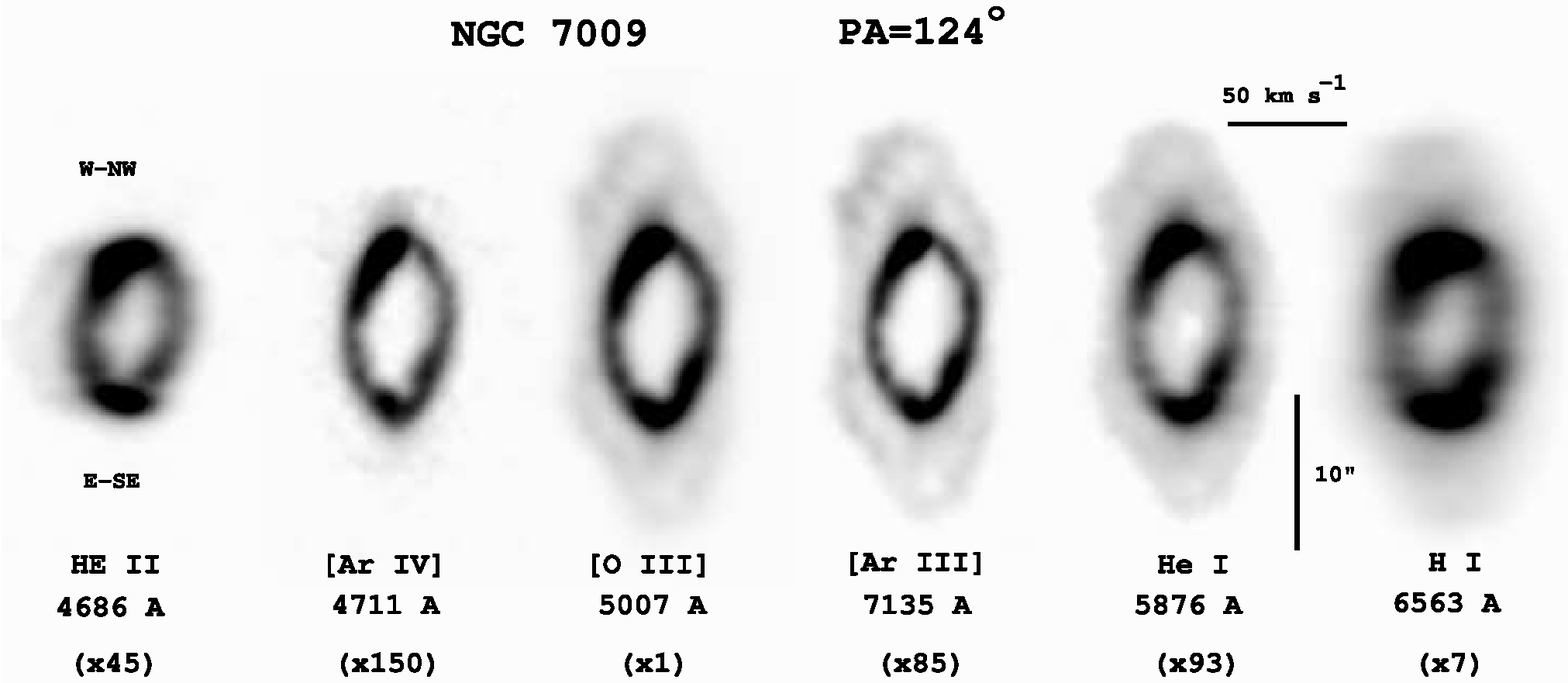}
   \caption{The NTT+EMMI spectral image of 
NGC 7009 at PA=124$\degr$ in six ionic species, in order - left-to-right - of decreasing IP. The slit 
orientation is superimposed to the left-most (He II) spectral image. The original flux is multiplied by the factor given in 
parenthesis, to make each emission graphically comparable with [O III]  $\lambda$5007 $\rm\AA\/$. More details 
are in the text.} 
\label{Figure 5}
\end{figure*}

\begin{figure}
   \centering \includegraphics[width=9cm,height=7cm]{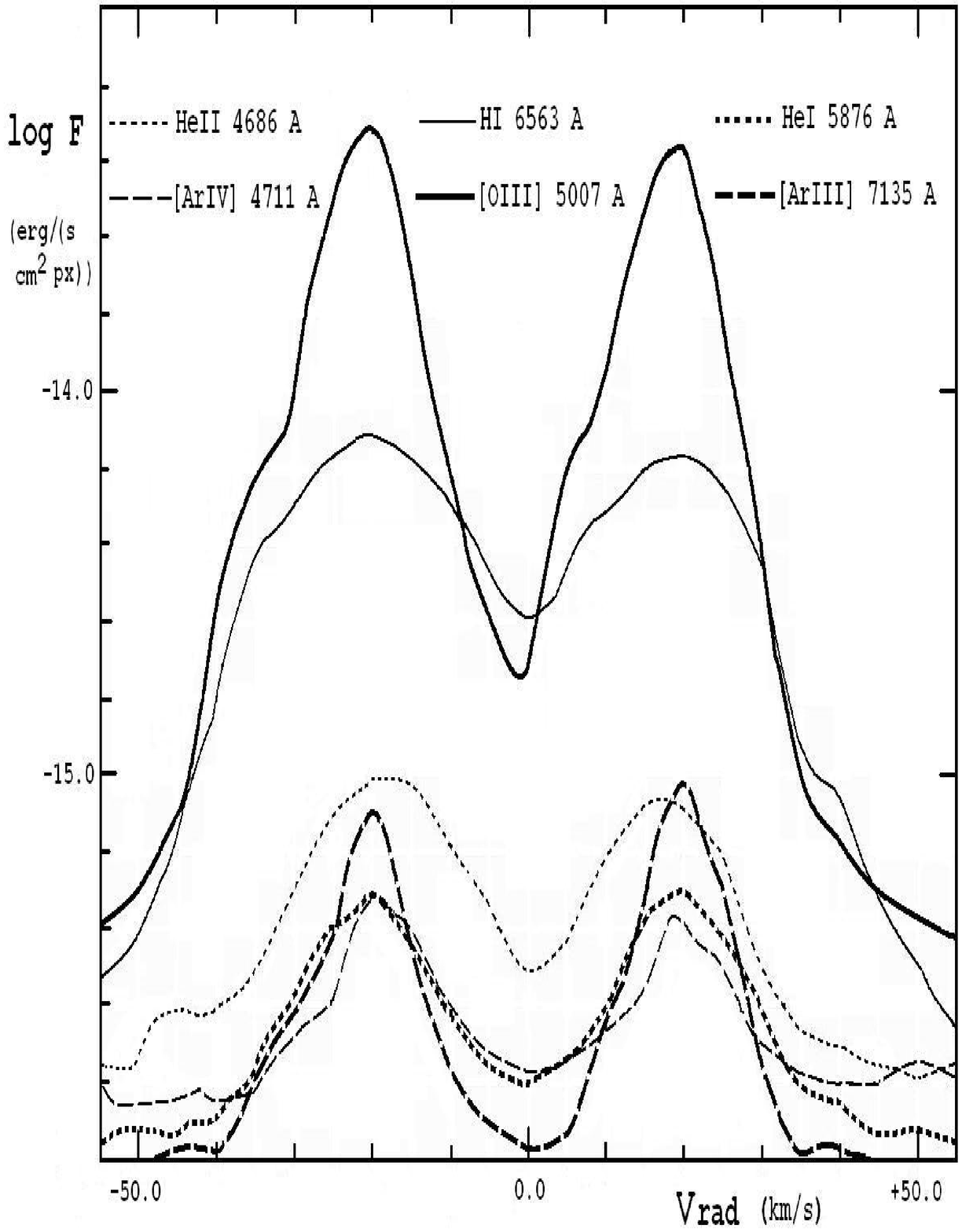}
   \caption{The logarithmic cspl flux-profile in main emissions of NGC 7009 at PA=124$\degr$. The observed (i. e., uncorrected for 
interstellar extinction) flux is in erg s$^{-1}$ cm$^{-2}$ px$^{-1}$. The X-axis is centered on the  
systemic velocity.}
\label{Figure 6}
\end{figure}

\paragraph{3.1.1.1.\,\,\,The cspl of the main shell}

\hfill

\hfill

The instrumental spectral resolution of our echellograms, estimated from  night-sky lines and comparison arc spectra, amounts to  
$\lambda$/$\Delta\lambda$=60\,000 (i. e., $W_{instr}$=5.0 km s$^{-1}$), and is common (to within 0.5 km s$^{-1}$) to all 
ionic species. Moreover, we refer to Sect. 2 for the fine-structure broadening ($W_{fs}$) in different recombination lines of hydrogen-like ions. 
\begin{figure}
  \centering \includegraphics[width=9cm,height=8cm]{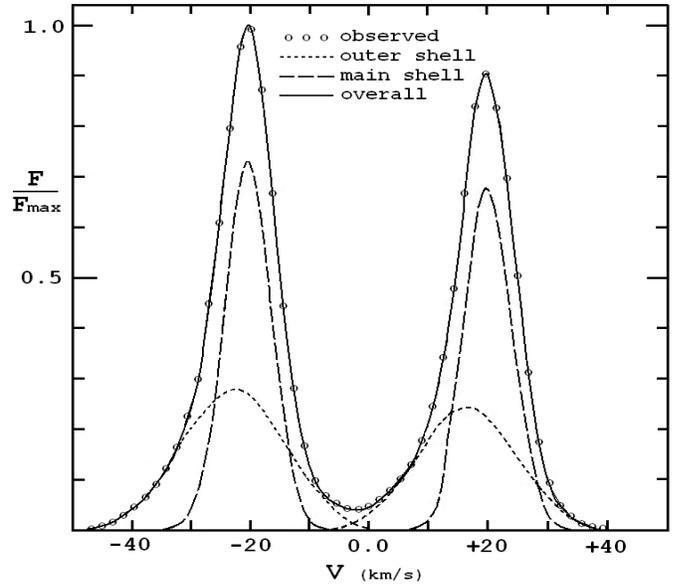}
   \caption{Multi-Gaussian analysis of the cspl velocity-profile. The relative [O III]  $\lambda$ 5007 $\rm\AA\/$ flux-profile observed in NGC 7009 
at PA=124$\degr$ (empty circles) is deconvolved in the 
main shell (long-dashed curve) and outer shell (short-dashed curve) contributions, and in their sum (continuous curve).}
\label{Figure 7}
\end{figure}
Other relevant parameters 
derived (or adopted) for the determination of turbulence, $W_{turb}$, in the cspl of the main shell of NGC 7009 at PA=124$\degr$ are 
listed in Table 1\label{Table 1}, where:
\begin{description}
\item[ - cols. 1 and 2] identify the emission line and the parent ion; 
\item[ - cols. 3 and 4] give the $W_{total}$ for blue- and red-shifted components of the main shell; 
\item[ - col. 5] contains the adopted value of $T_{\rm e}$ (according to 
\cite{hya95a} 1995a, 1995b; and \cite{sab04} 2004); 
\item[ - cols. 6 and 7] give the FWHM of E-SE and W-NW zvpc-peaks in the spectral image at PA=124$\degr$, providing - after correction 
for seeing+guiding - the 
radial thickness ($\Delta$r$\arcsec$) of each emitting layer 
in the main shell of NGC 7009.\footnote{In general, the value of $\Delta$r(cspl) 
can be inferred from the $\Delta$r(zvpc) trend at different PA. In the present case of 
NGC 7009, $\Delta$r(zvpc)  is nearly constant along and close to the apparent minor axis of the nebula 
(i. e., at PA=4$\degr$, 124$\degr$, 139$\degr$, 154$\degr$, and 169$\degr$).}  From $\Delta$r$\arcsec$(cspl)$\simeq$$\Delta$r$\arcsec$(zvpc),  
$V_{\rm exp}$(main shell)=4.0$\times$r$\arcsec$ km s$^{-1}$, 
and $\Delta$$V_{\rm exp}$(main shell)=$W_{grad}$=4$\times\Delta$r$\arcsec$ km s$^{-1}$, we have 
\begin{equation} 
W_{grad}=4\times[(\frac{(\Delta r_{(E-SE)} + \Delta r_{(W-NW)}}{2})^2 - W_{(s+g)}^2]^{0.5},
\end{equation}
where $W_{(s+g)}$  - the seeing+guiding contribution given by the FWHM of the central star's continuum - amounts to 
0.84($\pm$0.05)$\arcsec$ for all emissions; and 
\item[ - cols. 8 and 9] contain the resulting value of $W_{turb}$ for blue- and red-shifted components of the cspl profile (as given by Eq. (1)). 
They provide the 1-D overall turbulence map 
in the main shell of NGC 7009 through the $V_{\rm exp}$(km s$^{-1}$)=4.0$\times$r$\arcsec$ relation.
\end{description}

In spite of the large number of involved parameters, the estimated turbulence-accuracy in Table 1\label{Table 1} is $\pm$2 to $\pm$4 km s$^{-}1$ for
strong to faint collisionally-excited lines, $\pm$3 km s$^{-1}$ for He I $\lambda$ 5876 $\rm\AA\/$, 
and $\pm$4 km s$^{-1}$ for H I H$\alpha$ and He II $\lambda$ 4686 $\rm\AA\/$. 

\begin{centering}
\begin{table*}
\caption{The cspl-turbulence in the main shell of NGC 7009 at PA=124$\degr$. }
\begin{tabular}{ccccccccc}
\hline
\\
$\lambda$ &Ion &$W_{total}$(blue)&$W_{total}$(red)&$T_{\rm e}$&$\Delta$r$_{(E-SE)}$&$\Delta$r$_{(W-NW)}$&$W_{turb}$(blue)&$W_{turb}$(red)\\
($\rm\AA\/$)&&(km s$^{-1}$)&(km s$^{-1}$)&(K)&($\arcsec$)&($\arcsec$)&(km s$^{-1}$)&(km s$^{-1}$) \\
\\
\hline
\\
4686&He II&17.5&17.3&11\,000&1.73&1.84&3.6&2.5\\
4711&[Ar IV]&9.9&10.1&10\,500&1.55&1.65&5.5&5.8\\
5007&[O III]&10.0&10.0&10\,000&1.68&1.72&3.4&3.4\\
5876&He I&14.9&14.9&10\,000&1.70&1.70&6.2&6.2\\
6563&H I&26.1&25.5&10\,000&2.42&2.38&7.6&5.2\\
7135&[Ar III]&8.9&9.1&10\,000&1.50&1.46&4.2&4.6\\
\\
\hline
\end{tabular}
\end{table*}
\label{Table 1}
\end{centering}

\paragraph{3.1.1.2.\,\,\,The cspl of the outer shell}

\hfill

\hfill

The application of the foregoing procedure to the cspl of the outer shell in NGC 7009 at PA=124$\degr$ gives unreliable turbulence values due to the huge 
value (and large  
uncertainty) of $W_{grad}$, which is the predominant broadening component of the velocity profile. 

We consider $\lambda$ 5007 $\rm\AA\/$ of [O III] (Figs. 5, 6, and 7). On the one hand, we obtain $W_{total}$(blue)=21.3($\pm$0.3) km s$^{-1}$  
and $W_{total}$(red)=20.0($\pm$0.3) km s$^{-1}$. On the other hand, the [O III] thickness $\Delta$r(outer shell) - given by the zvpc profile corrected for 
seeing+guiding - amounts to 6.5($\pm$0.5)$\arcsec$; it provides $W_{grad}$=20.5($\pm$3.5) km s$^{-1}$ (through the $V_{\rm exp}$(km s$^{-1}$)= 
3.15($\pm$0.3)$\times$r$\arcsec$ relation valid in the outer shell). 

Thus, the mere acritical, and  slavish application of Eq. (1) to  [O III] blue- 
and red-shifted components of the cspl in the outer shell furnishes $W_{turb}$(blue)=-4.5($^{+15.5}_{-9.3}$) km s$^{-1}$, and 
$W_{turb}$(red)=-8.6($^{+16.9}_{-6.9}$) km s$^{-1}$. Even worse results are obtained from other, fainter emissions in Figs. 5 and 6. 

{\it All this stresses the leading role of the spatio-kinematical reconstruction, whose uncertainties mark actual limits of the turbulence-analysis in PNe} 
(to be noticed: in the absence of a spatio-kinematical recovery, the broad velocity profile in the outer shell of NGC 7009 could be ascribed to 
turbulent motions). 

In next sections, dedicated to the 2- and 3-D mapping of turbulence, we focus on the 
net spectral-profile of the main shell, ignoring the outer shell (and other sub-systems) of the Saturn Nebula.

\subsubsection{Ionic and overall 2-D turbulence-maps in the main shell of NGC 7009 at PA=124$\degr$} 
According to Sect. 2, the analysis of the velocity profile in all dlpl of a spectral image provides (through Eq. (1)) the 2-D turbulence-map of the 
parent ion within the entire nebular slice covered by the spectrograph slit. 

The [O III] results for the main shell of NGC 7009 at PA=124$\degr$ are 
graphically shown in Fig. 8\label{Figure 8}, where each dlpl is identified by its angular distance from the cspl (d$\arcsec$), in units of the zvpc ionic 
radius r(O$^{++}$)=5.55$\arcsec$. 
For practical reasons, Fig. 8\label{Figure 8} contains $W_{total}$(blue) and $W_{total}$(red) for dlpl at low-to-mean latitudes of the 
spectral image, whereas at high-latitudes (a) blue- and red-shifted components tend to merge,  
(b) spatial properties of the gas prevail on dynamical ones, and (c) the determination of $W_{total}$ becomes problematic. 
Thus, for all high-latitude dlpl (i. e., d/r(O$^{++}$)$\ge$0.85), we  are forced to measure the ``overall'' 
FWHM of the deconvolved spectral profile, $W_{over}$, corresponding to 

\begin{equation}
W_{over}=V_{\rm r}(red) - V_{\rm r}(blue) + (\frac{W_{total}(blue) + W_{total}(red)}{2}), 
\end{equation} 
where $V_{\rm r}$ is the radial velocity of red- and blue-shifted components in each dlpl. 
\begin{figure}
   \centering \includegraphics[width=9cm,height=8cm]{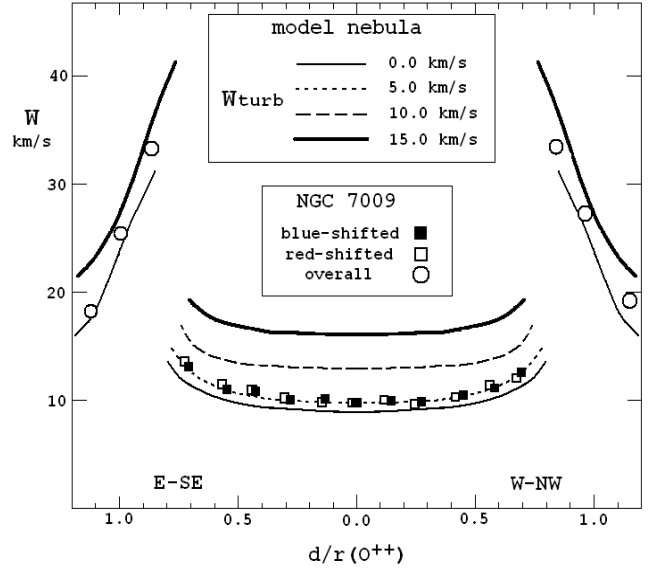}
   \caption{Main properties of dlpl in the [O III] main shell of NGC 7009 at PA=124$\degr$ (sampled along the slit with a  
three-pixel step, i. e., 0.84$\arcsec$), superimposed to the adopted model-nebula mimicing the spatio-kinematics of the main shell in the Saturn Nebula, and 
characterized by different values of W$_{turb}$. The x-axis 
contains the angular distance from the cspl (d), in units of the zvpc ionic radius r(O$^{++}$). At low-to-mean latitudes 
(i. e., d/r(O$^{++}$)$\le$0.85) W refers to   
W$_{total}$  for blue- and red-shifted components of each dlpl (squares), whereas at high-latitudes it provides the FWHM of the whole 
velocity-profile, i. e.,  W$_{over}$ (circles).}
\label{Figure 8}
\end{figure}

For comparison, Fig. 8\label{Figure 8} also shows corresponding results for the adopted model-nebula (assuming different values for the turbulence). 
This circular ring-model  \footnote{The circular assumption applies only to untilted spectral images at PA=124$\degr$. In the other observed PA 
of NGC 7009, the model-nebula is a tilted ellipse (see Fig. 6 in \cite{sab04} 2004).} mimics 
[O III] spatio-kinematical properties of the main shell of NGC 7009 at PA=124$\degr$ (i. e., r(O$^{++}$)=5.55$\arcsec$, 
$\Delta$r(O$^{++}$)$_{corr}$/r(O$^{++}$)=0.27,\footnote{Of course, $\Delta$r(O$^{++}$)$_{corr}$=($\Delta$r(O$^{++}$)$_{obs}^2$ - W$_{(s+g)}^2$)$^{0.5}$.}
$V_{\rm exp}$=4.0$\times$r$\arcsec$ km s$^{-1}$ and $T_{\rm e}$=10$^4$ K), and is observed with the same instrumentation and 
seeing+guiding conditions as NGC 7009. 

Using the $V_{\rm exp}$(main shell)=4.0$\times$r$\arcsec$ km s$^{-1}$ relation and simple geometrical considerations, we are able to map turbulent 
motions within the de-projected [O III] spectral image at PA=124$\degr$ 
(ionic 2-D turbulence-map), and the entire procedure can be repeated for  
other ionic species. The combination of all ionic 2-D tomographic-maps provides the overall 2-D tomographic-map of turbulence  
across the main shell of NGC 7009 
at PA=124$\degr$. This is shown in Fig. 9\label{Figure 9}, where: 
\begin{description}
\item [-] the H$\alpha$ contribution has been omitted (for obvious reasons); 
\item [-] for  He II we have used the 2-D turbulence map given by  $\lambda$ 6560 $\rm\AA\/$ (this line 
is fainter, but also a better turbulence diagnostic than $\lambda$ 4686 $\rm\AA\/$); 
\item [-] the 2-D turbulence reconstruction stops at high-latitudes (since the 
turbulence-accuracy decreases for d/r$\ge$0.85, due to the merging of blue- and red-shifted components); and   
\item [-] 
pixelization is due to the discrete dlpl-sampling  in each spectral image, and fuzziness to the partial 
superposition of turbulence-values in different ionic species.
\end{description}
\begin{figure}
   \centering \includegraphics[width=9cm]{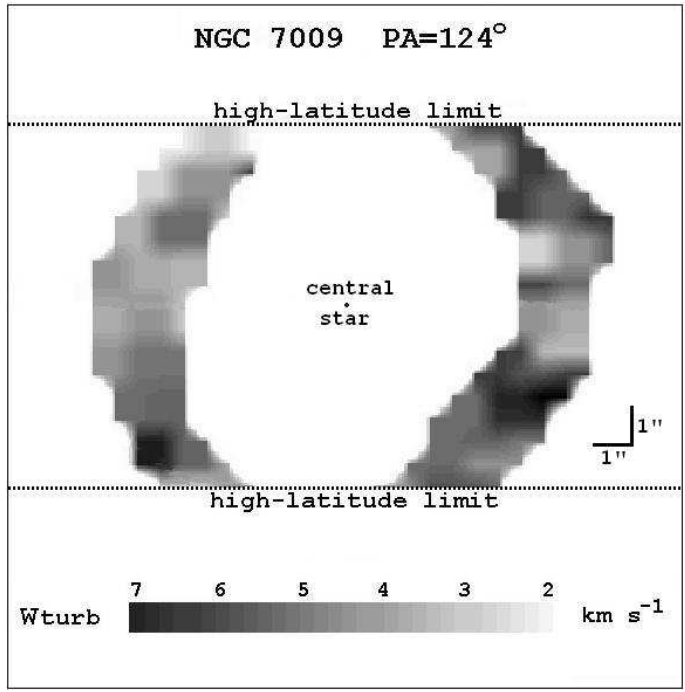}
   \caption{The preliminary (i. e., under-sampled) tomographic  turbulence-map  in the main shell of NGC 7009 at PA=124$\degr$, obtained 
by assembling  low-to-mean 
latitude 2-D turbulence maps in [O III], [Ar III], [Ar IV], He I and He II ($\lambda$ 6560 $\rm\AA\/$). Same orientation as Fig. 5.}
\label{Figure 9}
\end{figure}

\subsection{The overall 3-D turbulence-map in the main shell of NGC 7009}

The application of the whole procedure described in the foregoing section to all observed PA of NGC 7009  extracts the  
overall 2-D turbulence-map in twelve, equally spaced radial slices of the nebula covered by the spectrograph slit.  
Their assembly 
by means of our 3-D rendering program  (\cite{rag01} 2001), gives the spatial distribution of the gas-turbulence within the (almost) 
complete main shell (overall 3-D turbulence-map).  

The preliminary (i. e., under-sampled) overall 3-D turbulence-map in the main shell of the Saturn Nebula is shown in Fig. 10 and in MOVIE N. 1 
(on-line material), where: 
\begin {description} 
\item [ - ] the $W_{turb}$-range is 2.0 km s$^{-1}$ (faintest regions) to  8.0 km s$^{-1}$ (brightest regions); 
\item [ - ] the left-most (00,00) panel in Fig. 10, and the fixed image at the start of the movie represent the nebula as seen from the Earth 
(North is up, and East to the left); and 
\item [ - ] the irregular, vertical, empty-band visible at (00,90) identifies high-latitude dlpl (i. e., d/r$\ge$0.85) in original spectral images.
\end{description}

\begin{figure*}
   \centering \includegraphics[width=18cm,height=4cm]{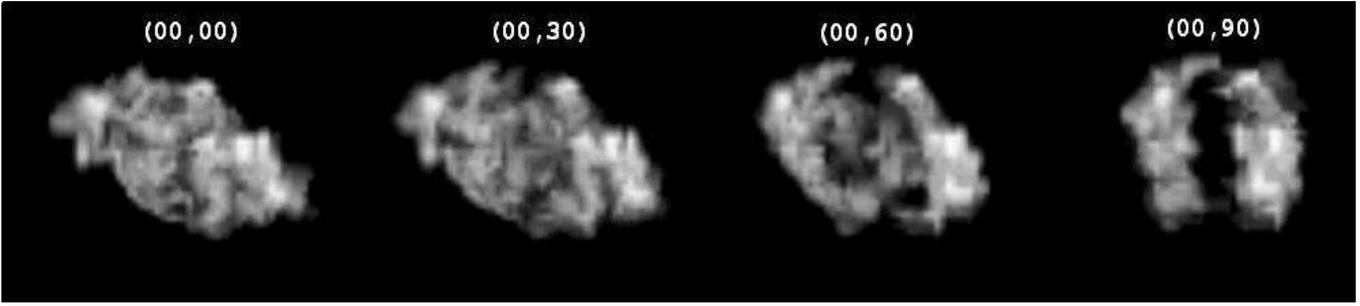}
   \caption{Snapshots of the turbulence-structure in the main shell of NGC 7009 for a rotation around the North-South axis 
centered on the exciting star.  The $W_{turb}$-range is 2.0 km s$^{-1}$ (faintest regions) to  8.0 km s$^{-1}$ (brightest regions). The line of sight 
is given by ($\theta,\psi$), where $\theta$ is the zenith angle, and $\psi$ the
azimuthal angle. 
The left-most (00,00) image corresponds to the nebula as 
seen from the Earth (North is up and East to the left). The complete 3-D turbulence-movie is shown in the electronic version of the paper 
(on-line data).
}
\label{Figure 10}
\end{figure*}

Although the detailed recovery and analysis of the  1-, 2-, and 3-D turbulence in NGC 7009 are beyond aims of this section (they 
are deferred to a dedicated paper, in preparation), we wish to underline the patchy turbulence-distribution within the main shell, and modest 
turbulence-values of innermost nebular regions at 
the highest ionization degree (i. e. He II and Ar IV emissions). 
In particular, the latter result 
\begin{description}
\item[-] is surprising for an extended X-ray source 
(\cite{gue02} 2002) 
like NGC 7009, whose hot and luminous powering-star (log T$_*$(K)=4.95, 
log L$_*$/L$_{\odot}$ =3.70; 
\cite{sab04} 2004) 
is ejecting fast matter at a high - although uncertain - rate (terminal velocity$\simeq$2700 km s$^{-1}$, and mass-loss in the range 
3.0$\times$10$^{-10}$ to 1.0$\times$10$^{-8}$ M$_{\odot}$ yr$^{-1}$; 
\cite{cer85} 1985, 1989; \cite{bom86} 1986; \cite{hut89} 1989; \cite{tin02} 2002); 
\item[-] contrasts with  hydrodynamical expectations 
for shocked nebular regions interacting with the hot-bubble generated by  fast stellar winds 
(\cite{fem94} 1994; \cite{mel97} 1997; \cite{dwa98} 1998; \cite{gar99} 1999); and 
\item[-] needs confirmation by the spatio-kinematical analysis of forbidden lines at an even higher ionization degree 
(e. g., $\lambda$ 7005 $\rm\AA\/$ 
of [Ar V], IP=59.8 eV, and $\lambda$ 3425 $\rm\AA\/$ of [Ne V], IP=97.1 eV ). 
\end{description}

\section{Discussion }
Current theoretical models and radiation-hydrodynamical simulations of PNe predict that random, unsteady, non-thermal motions at a local scale constitute a 
fundamental characteristic of the expanding gas. 

Actually, our present knowledge on the turbulence-structure of real PNe is nebulous, since (1) this elusive parameter 
has modest kinematical effects on the complex velocity profile, and (2) practical determinations  - 
providing a unique, average turbulence-value for 
each PN (\cite{gue98} 1998; \cite{ack02} 2002; 
\cite{gaz03} 2003, 2006; \cite{gz03} 2003, 2007; and  \cite{med06} 2006) - are affected by one or more weakenesses, namely: 
\begin{description}
\item[(i)] poor observational material (e. g., insufficient spectral resolution, very partial coverage of the nebula, limited number of emissions); and/or 
\item[(ii)] coarse data reduction (integration along the slit, assumption of spherical symmetry); and/or  
\item[(iii)] inadequate analysis (omission of some important  
broadening factor, absence of a  spatio-kinematical reconstruction). 
\end{description}

For these reasons, we have carried out a detailed analysis based on the three-step methodology (spatio-kinematics, tomography and 3-D recovery) 
developed at the Astronomical Observatory of Padua. The disentanglement of all broadening components  
in the velocity profile (instrumental resolution, 
thermal motions, turbulence, expansion gradient, and fine-structure 
of hydrogen-like ions) allows us the extraction of disordered non-thermal  motions, and their mapping in 
1-, 2-, and 3-dimensions. 

Although the observed velocity-structure of a PN depends on several factors - 
since $W_{grad}$=f(ionic distribution + expansion law), $W_{instr}$=f(instrumental set-up), 
$W_{ther}$=f(electron temperature), $W_{turb}$=f(turbulence), and $W_{fs}$=f(atomic properties) -, we argue that 
the final turbulence-accuracy  mainly depends on the reliability of its spatio-kinematical reconstruction. 

Thus, 
{\it the detailed study of turbulence, of its stratification within each target and  (possible) systematic variation among 
different sub-classes  of PNe (A)  is strictly connected to the general problem of de-projecting  the bi-dimensional, 
apparent morphology of a three-dimensional mass of gas; and  (B) needs (and deserves) long-slit, high-spectral resolution, multi-PA echellograms 
over a wide spectral range containing emissions at low-, mean-, and high-ionization, to be reduced and analyzed 
with a straightforward and versative methodology extracting the whole physical information stored in each 
frame at best. }

We underline that {\it the foregoing paradigm} - implying the recovery and pixel-to-pixel 
analysis of a large sample of emission lines in an adequate number of PA - {\it should be extended to all observational parameters 
(e. g., flux, electron temperature, electron density, ionic and chemical abundances, 
etc.), due to the spatio-kinematical florilegium exhibited by PNe} (\cite{tur02} 2002; \cite{ben03} 2003; \cite{sab04} 2000, 2004, 2005, 2006). 

We end  
pointing out that the methodology developed here for PNe actually applies to all 
types of expanding nebulae (including nova and supernova 
remnants, shells around Population I Wolf-Rayet stars, 
nebulae ejected by symbiotic stars, bubbles surrounding early spectral-type main sequence stars, etc.) covered at adequate ``relative 
spatial''  and 
``relative spectral'' resolutions.
\section{Conclusions}
So far, the quantification of stratified turbulent motions in the ionized gas of PNe - as predicted by hydrodynamical 
simulations and theoretical models - escaped a direct verification because of its strict connection with the general, long-standing astronomical problem of 
de-projecting the bi-dimensional, 
apparent morphology of a three-dimensional mass of gas. The solution is now approached by spatio-kinematical, tomographic, and 3-D rendering analyses applied to 
long-slit, multi-PA, high-resolution  echellograms over a wide spectral range. 

We develop an accurate
procedure disentangling all broadening components (instrumental resolution, thermal motions, turbulence, expansion 
gradient, and fine structure of hydrogen-like ions) of the velocity profile. This overcomes  main  weaknesses 
affecting  previous  determinations, and allows us to quantify and map in 1-, 2-, and 3-dimensions the gas-turbulence in PNe. 

The multi-dimensional turbulence-analysis - soon applied to a representative sample of targets in different evolutionary phases, and  
combined with 
expectations from updated radiation-hydrodynamical simulations -  will provide new, deep insights on different 
shaping agents of PNe (ionization, wind interaction, magnetic fields, binarity of the central star, etc.), and on their mutual interaction.  

Last, we emphasize all advantages - and encourage the adoption - of long-slit  echellograms 
covering the entire instrumental spectral range, since the usual insertion of an 
interference filter isolating a single echelle 
order proves superfluous and strongly limitative in most cases... it is like driving a Ferrari with the hand-brake applied.

\begin{acknowledgements} 
This paper is dedicated to the memory of Prof. Mario Perinotto, an expert and generous colleague, an example of intellectual 
honesty, and, mainly,  a dear friend... ciao, Mario.



This work was supported by grant no. 2006022731 of the PRIN of Italian Ministry of University and Science Research.

\end{acknowledgements}


\end{document}